\documentclass[11pt]{article}

\usepackage{amsmath}
\usepackage{epsfig}
\usepackage{graphics}

\newcommand{\N}{N\raise.7ex\hbox{\underline{$\circ $}}$\;$}

\begin{document}

\title{
E.M. Ovsiyuk\footnote{Mozyr State Pedagogical University named after I.P. Shamyakin, Belarus; e.ovsiyuk@mail.ru}\\[4mm]
  Quasi-plane waves for a particle with spin 1/2  on the background of
  Lobachevsky
 geometry:  simulating of a special medium}

\maketitle

\begin{abstract}

In the paper complete systems of exact solutions for Dirac and Weyl equations in the Lobachevsky space $H_{3}$ are constructed on the base of the method of separation of the variables in quasi-cartesian coordinates. An extended helicity operator is introduced. It is shown that solution constructed when translating to the limit of vanishing curvature coincide with common plane wave solutions on Minkowski space going in opposite $z$-directions. It is shown the problem posed in Lobachevsky space simulates a situation in the flat space for a quantum-mechanical particle of spin 1/2 in a 2-dimensional potential barrier smoothly rising to infinity on the right.

\end{abstract}


 \hspace{5mm} It is known that in the field theory of elementary
particles,  the basis of plane wave  is of the  most
 use. However, in presence of a  curvature, any common plane wave solutions  do
  not exist. Therefore, of a  special interest are
examples non-Euclidean spaces in which some  analogues of such
solutions can be constructed.  In the paper \cite{Shapiro-1962},
it was shown that in the Lobachevsky space there are such
solutions for particles with spin 0; also see the
 books by Gelfand--Graev--Vilenkin \cite{Vilenkin-Smorodinsky-1964},
\cite{Gelfand-Graev-Vilenkin-1962}. An  analog of plane waves in a
space of constant positive curvature was studied by Volobuev
\cite{Volobuyev}.
   The later treatment of this problem was
    given in \cite{Ovsiyuk-Tokarevskaya-Red'kov-NPCS-2009}.
Solutions of the  plane wave type for Maxwell's equations have
been considered in
\cite{Bychkovskaya-Vesti-2006}--\cite{Ovsiyuk-Red'kov-Vesti-2009}.
In \cite{Kurochkin-Otchik-2011},   the problem of constructing
solutions of the Dirac equation in the Lobachevsky space was
studied  on the  base of the method of squaring;  in particular, it was
pointed  out  the possibility of constructing  solutions of
the  Dirac plane wave studied  starting with Shapiro's  scalar waves .
In this paper we will construct a complete basis of solutions of the
plane wave type  for Dirac and Weyl particles in the Lobachevsky  space,
applying  the method of separation of the
variables in a special system of quasi-cartesian coordinates closely
related to horospherical coordinates.

To understand the physical meaning  of the system under consideration, it  should be mentioned
that
Lobachewsky geometry simulates a  medium with special consti\-tutive relations.
The situation is specified in quasi-cartesian coordinates $(x,y,z)$ was treated in \cite{Ovsiyuk-Red'kov-Vesti-2009}.
Exact solutions of the Maxwell equations in  complex  3-vector  form, extended to curved
space models within the tetrad formalism, have been found in Lobachevsky space.
The problem reduces to a second order diffe\-rential equation which can be associated with an 1-dimensional
Schr\"{o}dinger problem for a particle in external  potential field $U(z) =  U_{0}e^{2z}$.
In quantum mechanics, curved geometry acts as an effective potential barrier with reflection coefficient $R=1$;
in electrodynamic  context  results similar to quan\-tum-mechanical  ones   arise:
the Lobachevsky  geometry   simu\-lates  a  medium  that effectively  acts as
an ideal  mirror.
Penetration of the electromagnetic field into the effective medium, depends on the parameters of an
electromagnetic wave,
frequency $\omega, \; k_{1}^{2} + k_{2}^{2}$, and the curvature radius $\rho$.

In the present paper, that analysis will be extended to
 the case of particles with spin $ 1/2$, described by equations of Dirac and Weyl.
The  generalized spinor  plane waves   can find
application in the analysis of the behavior  of fermions particles
on cosmological scales, or in simulating special media affecting the spinor particles.

\subsection*{1 On the solutions of the Schr\"{o}dinger equation}

 \hspace{5mm} In the Lobachevsky space--time parameterized by
 quasi-cartesian coordinates
$$
dS^{2}= dt^{2} - e^{-2z} (
dx^{2} + dy^{2} ) - dz^{2} \; ;
$$

\noindent   the element of volume is given by
$$
 dV =\sqrt{-g} \; dx dy dz = e^{-2z} dx dy dz \;, \qquad  x,y,z \in ( - \infty , + \infty )\; .
$$

\noindent    The magnitude and sign of the  $z$   are substantial,
in particular when referring to the probabilistic interpretation of  the wave functions
$$
dW = \mid \Psi \mid^{2} d V =  \mid \Psi \mid^{2} e^{-2z}  \; dx dy dz\,.
$$

Let us describe some details of the parametrization of the space  by coordinates $(x,y,z)$.
It is known that this model can be identified with a branch of hyperboloid in 4-dimension flat space
$$
u_{0}^{2} - u_{1}^{2} - u_{2}^{2} - u_{3}^{2} = \rho^{2} \; , \qquad
u_{0} = + \sqrt{\rho^{2}  + {\bf u}^{2} } \;  .
$$

Coordinates   $x,y,z$ are referred to   $u_{a}$ by relations
$$
u_{1}=x e^{-z}\; , \;\; u_{2}=y e^{-z}\; , \;\;
$$
$$
u_{3}={1\over 2}[(e^{z}-e^{-z})+(x^{2}+y^{2})e^{-z}]\; ,
$$
$$
u_{0}={1\over 2}[(e^{z}+e^{-z})+(x^{2}+y^{2})e^{-z}]\; .
\eqno(1.1a)
$$

\noindent It is convenient to employ 3-dimensional Poincar\'{e} realization for
Lobachevsky  space as  an inside part of 3-sphere:
$$
q_{i} = {u_{i} \over  u_{0}}
 = {u_{i} \over  \sqrt{\rho^{2} + u_{1}^{2} + u_{2}^{2} + u_{3}^{2} }}, \qquad q_{i}q_{i} < +1 \; .
\eqno(1.1b)
$$
Quasi-cartesian coordinates $(x,y,z)$  are referred to  $q_{i}$ as follows
$$
q_{1} = {2 x \over  x^{2} + y^{2}  + e^{2z} + 1}\;, \qquad
q_{2} = {2 y \over  x^{2} + y^{2}  + e^{2z} + 1}\; ,
$$
$$
q_{3} = {x^{2} + y^{2} + e^{2z} -1 \over z^{2} + y^{2}  + e^{2z} +
1} \,.
\eqno(1.1c)
$$

\noindent Inverses to $(1.1c)$ relations are
$$
x =  {q_{1} \over 1 - q_{3}} \; , \qquad y =    {q_{2} \over 1 -
q_{3}} \; , \qquad e^{z} =  { \sqrt{1 - q^{2}} \over 1 - q_{3}}
\; .
\eqno(1.1d)
$$

In particular, note that on the  axis $q_{1}=0, q_{2}=0, q \in (-1 , +1)$ relations $(1.1d)$
assume the form
$$
x =  0 \; , \qquad y = 0 \; , \qquad e^{z} =   \sqrt{ {1 + q_{3}
\over 1 - q_{3}}  }    \,,
$$

\noindent
that is
$$
q_{3} \longrightarrow  +1 \; , \qquad e^{z} \longrightarrow + \infty \; , \qquad z \longrightarrow  + \infty \; ;
$$
$$
q _{3} \longrightarrow  -1 \; , \qquad e^{z} \longrightarrow  + 0 \; ,
\qquad  z \longrightarrow - \infty \,.\eqno(1.2)
$$

Schr\"{o}dinger equation in Riemannian space \cite{Red'kov-book-2009}
$$
 i   \, \hbar\,\partial  _{t }      \Psi
=
 {1 \over 2M} \left [  ({i \,\hbar  \over \sqrt{-g} }
\partial_{k} \sqrt{-g}  + e   A_{k} ) (- g^{kl}
)(i \,\hbar\, \partial  _{l } +  e A_{l})  \right ]
    \Psi \,
$$

\noindent in quasi-cartesian coordinates $ (1.1a) $
takes the form
$$
i\, \hbar\,{\partial \over \partial t}\Psi=-{ \hbar^{2} \over 2M}\,
\left[e^{2z}\,{\partial^{2} \over \partial x^{2}}+
e^{2z}\,{\partial^{2} \over \partial y^{2}}+e^{2z}\,{\partial\over \partial z}\,e^{-2z}\,
{\partial\over \partial z}\right]\Psi\,.
$$

\noindent
The variables are separated by the substitution
 $ \Psi= e^{-iEt /\hbar} e^{ik_{1}x}\,e^{ik_{2}y}\,f(z)$:
$$
\left[{d^{2}\over dz^{2}}-2\,{d\over dz} + \epsilon -e^{2z}(k_{1}^{2}+k_{2}^{2}) \right]\,f(z)=0 \; ,
\eqno(1.3a)
$$

\noindent where a dimensionless quantity used $\epsilon= 2ME\rho^{2}/\hbar^{2}$,\; $\rho$ --
curvature  radius of the  space.
Elementary substitution $f= e^{z} \varphi (z)$ in  equation $(1.3a)$ gives a
 Schr\"{o}dinger-like  equation
$$
\left ( { d^{2} \over dz^{2} }   +  \epsilon -1  - (k_{1}^{2} + k_{2}^{2}) e^{2z} \right )  \varphi (z) = 0
\eqno(1.3b)
$$

\noindent with potential function
$$
U(z) = 1  + (k_{1}^{2} + k_{2}^{2}) e^{2z} \,. \qquad
\eqno(1.3c)
$$

\noindent
Note that the probabilistic interpretation of the wave function
after the transformation  to $\varphi$  reads
$$
dW = \mid \Psi \mid^{2} d V =  \mid \varphi \mid^{2}   \; dx dy dz \; .
\eqno(1.4)
$$

An  easily interpretable physical solution for $\epsilon>1$ is the following:
 on the left we have  the superposition of two waves, falling from the left and reflected.
On the right behind the barrier, the wave function must sharply decrease to zero.


It should be noted that the case $k_{1}=0,\; k_{2}=0$ is special:
 the equation $(1.3a)$ is very much changed  because the potential function  disappears
$$
\left ( {d^{2}\over dz^{2}}-2\,{d\over dz} + \epsilon  \right )\,f(z)=0 \; , \qquad
f = e^{(1 \pm i \sqrt{\epsilon -1}) z}\; , \qquad \varphi =  e^{( \pm i \sqrt{\epsilon -1}) z} \,,
\eqno(1.5)
$$

\noindent and the function $\varphi$   is a solutions of the type of ordinary plane wave.

Let us turn to the general case and in eq. $(1.3a)$  introduce  the variable
$$
\sqrt{k_{1}^{2} + k_{2}^{2}} \; e^{z} = Z\; , \qquad Z \in  (0 , + \infty ) \; ;
$$

\noindent  the equation takes the form
$$
\left ( {d^{2} \over dZ^{2} } - {1 \over   Z} {d \over d Z}  +{\epsilon \over  Z^{2}}  - 1 \right ) f (Z) = 0 \; ;
\eqno(1.6)
$$

\noindent
with the help of a substitution  $ f = \sqrt {Z} \; F $, one can remove the term with the first derivative
$$
\left ( {d^{2} \over dZ^{2} } +{ \epsilon  -3/4 \over  Z^{2}}  - 1 \right ) F  (Z) = 0 \; .
$$

\noindent This  form  makes it easy to find the asymptotical behavior  of solutions

\vspace{3mm}
$
\underline{(z \longrightarrow - \infty ) \;\; Z \rightarrow 0 \,,}
$
$$
F \sim Z^{1/2  \pm i \sqrt{\epsilon -1}} \;,
\qquad f \sim  Z^{1\pm i \sqrt{\epsilon -1}} \; , \qquad \varphi \sim  e^{\pm i \sqrt{\epsilon -1}\; z } \; ;
$$

$
\underline{(z \longrightarrow + \infty )\;\; Z \rightarrow + \infty \; ,}
$
$$
F \sim e^{\pm Z} \; , \qquad f = \sqrt{Z} \; e^{\pm Z} \;, \qquad \varphi \sim  e^{-z/2} \exp \left [ \pm \sqrt{k_{1}^{2} + k_{2}^{2}} \; e^{z} \right ]  .
$$
$$
\eqno(1.7)
$$

We now turn to the construction of exact solutions of (1.6) in the entire range of coordinate $z$.
We seek solutions in the form of $f (Z) = Z^{A} e^{BZ}\,F (Z)$; equation (1.6) gives
$$
Z\,{d^{2}F\over dZ^{2}}+(2A-1+2BZ)\,{dF\over dZ}+
$$
$$
+ \left ( (B^{2}-1)\,Z-B\,(1-2A)+{A\,(A-2)+\epsilon\over Z} \right )F=0\,.
\eqno(1.8)
$$

\noindent
At $ A, \, B $ chosen according (for definiteness, we take  the minus sign before
 the root in the expression for
$A$;  assuming $\epsilon >1$)
$$
A=1 - i \sqrt{\epsilon -1 }\,,\qquad B^{2} = 1\,,\eqno(1.9)
$$

\noindent the equation (1.8) is simplified
$$
Z\,{d^{2}F\over dZ^{2}}+(2A-1+2BZ)\,{dF\over dZ}-B\,(1-2A)\,F=0\,.\eqno(1.10)
$$

\noindent
In  (1.10), let us  make another change $ Z=y/2$:
$$
y\,{\,d^{2}F\over dy^{2}}+(2A-1+By)\,{\,d\,F\over dy} + B\,(A-{1\over2})\,F=0\,,\eqno(1.11)
$$

\noindent
with $ B =- 1 $  it is an equation for the confluent
hypergeometric function
$$
y\,{\,d^{2} Y \over dZ^{2}}+(c -y) {\,d\,Y \over dy}- a Y =0\, , \qquad
$$
$$
c =2a\;, \qquad a  = A- 1 / 2  =  1/2  -  i \sqrt{\epsilon -1}   \,,
$$
$$
f(Z)=y^{a+1/2}e^{-y/2}\,Y(y)\,.
\eqno(1.12)
$$

We use two pairs of linearly independent solutions\cite{Beytmen-1973}
$$
Y_{1} = \Phi (a,2a,y) \; , \qquad Y_{2} =  y^{1-2a} \Phi (1 - a, 2-2a, y)
$$
and
$$
Y_{5} = \Psi (a,2a,y) \; , \qquad  Y_{7} = e^{y} \Psi (a,2a,-y) \;,
\eqno(1.13)
$$

These pairs of solutions are related by Kummer linear relations  \cite{Beytmen-1973}
$$
Y_{5} = {\Gamma (1-2a) \over \Gamma (1-a) } \; Y_{1} + {\Gamma(2a-1) \over \Gamma (a)} \; Y_{2} \; ,
$$
$$
Y_{7} = {\Gamma (1-2a) \over \Gamma (1-a) } \; Y_{1} - {\Gamma(2a-1) \over \Gamma (a)} \; Y_{2} \; ,
\eqno(1.14a)
$$

\noindent
which after multiplication by $ y ^ {a +1 / 2} e ^ {-y / 2} $ take the form
$$
f_{5} = {\Gamma (1-2a) \over \Gamma (1-a) } \; f_{1} + {\Gamma(2a-1) \over \Gamma (a)} \; f_{2} \; ,
$$
$$
f_{7} = {\Gamma (1-2a) \over \Gamma (1-a) } \; f_{1} - {\Gamma(2a-1) \over \Gamma (a)} \; f_{2} \; .
\eqno(1.14b)
$$

Note that the solutions $ Y_ {1} $ and $ Y_ {2} $ describe the case of negative $ z \longrightarrow - \infty $
wave with the asymptotic behavior

\vspace{2mm}
\underline{$z \rightarrow - \infty \;, \; (y \longrightarrow 0 )$}
$$
f_{1} \sim  y^{a+1/2} = \left ( 2 \sqrt {k_{1}^{2} + k_{2}^{2}} \right ) ^{1 - i\sqrt{\epsilon -1}}\;
  e^{z}\;  e^{-i\sqrt{\epsilon -1}\;  z} \; ,
$$
$$
f_{2} \sim   y^{a+1/2} \; y^{1-2a} =  \left ( 2 \sqrt {k_{1}^{2} + k_{2}^{2}} \right )
^{1 + i\sqrt{\epsilon -1}}\;  e^{z} \; e^{+ i\sqrt{\epsilon -1}\;  z} \; .
\eqno(1.15)
$$

Thus, for example, the function $ Y_ {5} $ (and the related $ \varphi_ {5} $)
at negative $ z \longrightarrow - \infty $ behaves as a superposition of two plane waves according to
$$
\varphi_{5} \sim {\Gamma (1-2a) \over \Gamma (1-a) }
\left ( 2 \sqrt {k_{1}^{2} + k_{2}^{2}} \right ) ^{1 - i\sqrt{\epsilon -1}} e^{-i\sqrt{\epsilon -1}\; z }
 +
 $$
 $$
 +  {\Gamma(2a-1) \over \Gamma (a)}   \left ( 2 \sqrt {k_{1}^{2} + k_{2}^{2}} \right )
  ^{1 + i\sqrt{\epsilon -1}} e^{+i\sqrt{\epsilon -1}\; z} \; .
\eqno(1.16)
$$

We define  the reflection coefficient as  the square modulus of the amplitude ratio in a superposition of plane waves
$$
M_{-} \; e^{-i\sqrt{\epsilon -1}\; z } + M_{+} \; e^{+i\sqrt{\epsilon -1}\; z }\;, \qquad
R =  \left |  {M_{-} \over  M_{+}} \right | ^{2} ,
$$
$$
R =  \left | {\Gamma (1-2a) \over \Gamma (2a-1) } {\Gamma (a) \over  \Gamma (1-a) } \right |^{2} \; .
\eqno(1.17a)
$$

\noindent
We take into account
$$
1-2a = +2i\sqrt{\epsilon -1}\; , \qquad 2a-1 =  -2i \sqrt{\epsilon +1}\;,
$$
$$
a = 1/2  -  i \sqrt{\epsilon -1} \;, \qquad 1- a =  1/2  +  i \sqrt{\epsilon -1}\,,
$$

\noindent
then
$$
R =  \left | {\Gamma (+2i\sqrt{\epsilon -1}) \over \Gamma (-2i \sqrt{\epsilon -1}) }\right |^{2}
 \; \left |
{\Gamma (1/2  -  i \sqrt{\epsilon -1}) \over  \Gamma (1/2  +  i \sqrt{\epsilon -1}) } \right |^{2} \equiv 1 \; .
\eqno(1.17b)
$$

We find the behavior of $ Y_ {5} $ at large $ y $. Using the known asymptotic relation \cite{Beytmen-1973}
 $$
Y_{5} =  \Psi (a,c,y) \sim y^{-a}\,,
$$

 \noindent we get
$$
z \rightarrow + \infty \;, \qquad
f_{5} = y^{a+1/2}e^{-y/2}\,Y_{5}  \sim
 y^{1/2}e^{-y/2} \sim
 $$
 $$
 \sim
 \left ( 2 \sqrt{k_{1}^{2} + k_{2}^{2}} \; e^{z} \right )^{1/2}\;
 \exp \left ( - \sqrt{k_{1}^{2} + k_{2}^{2}} \; e^{z}\right  ) \longrightarrow \; \exp^{- e^{+\infty}}  = 0 \; .
\eqno(1.18)
$$

Thus, the solution $ f_ {5} $    describes  the expected
situation: wave going from the left is reflected with probability $ 1 $ on the effective barrier;
 behind the barrier  the solutions  sharply decrease  to zero.
 It is easy to find the critical point, after which
  wave function must sharply decrease
 $$
 \epsilon -1 = (k_{1}^{2} + k_{2}^{2} ) \; e^{2z} \qquad
 \Longrightarrow \qquad z_{0} = \ln  \sqrt{{ \epsilon -1 \over k_{1}^{2} + k_{2}^{2}}} \; ,
\eqno(1.19a)
$$

\noindent in the usual units, this critical point is described by the relation
$$
z_{0} =  \rho \; \ln  \sqrt{{ 2mE \rho^{2} /  \hbar^{2}  -1 \over (K_{1}^{2} + K_{2}^{2}) \rho^{2} }} \; .
\eqno(1.19b)
$$

The solution $ f_ {7} $ for $ y \longrightarrow + \infty $ goes to infinity
$$
f_{7} \sim y^{a+1/2} e^{-y/2} \; e^{y} y^{-a}  = y^{+1/2}   e^{+y/2} \; .
$$

\noindent
However, it is of no  clear physical interpretation.

However, it is easy to interpret such a solution $ f_ {1} $.
 Indeed, far on the right behind the barrier,
 taking into account the asymptotic formula  \cite{Beytmen-1973}
$$
y \rightarrow + \infty \;, \qquad \Phi (A,C,y) = {\Gamma (C) \over \Gamma (A)}e^{y} y^{A-C} \,,
$$

\noindent
we get
$$
z \longrightarrow + \infty\;, \qquad
f_{1} \sim y^{1/2} e^{y/2} \longrightarrow + \infty^{\infty} \; ;
\eqno(1.20)
$$

\noindent that is, for $ z \rightarrow + \infty $ is a real function with infinite probability density.
Far left is a plane wave propagating to the left.

Next we consider the analogue of this situation for particles with spin $1/2$,
described by the relativistic Dirac equation, when  analysis is much more complicated.

\subsection*{2 The Dirac equation in the space $ H_ {3} $, separation of variables}

\hspace{5mm}We start with the general covariant form of the Dirac
equation \cite{Red'kov-book-2009}
$$
\left [ i\; \gamma^{a} \; \left ( e_{(a)}^{\alpha}\;{\partial
\over \partial x^{\alpha}} + {1\over 2}\; ({1 \over
\sqrt{-g}}{\partial \over \partial x^{\alpha}} \sqrt{-g} \;
e_{(a)}^{\alpha}    ) \right )   \; - \; m \; \right  ] \;\Psi (x)
= 0 \; . \eqno(2.1)
$$

\noindent In the coordinate system (1.1) we use the tetrad
$$
e_{(a)}^{\beta} = \left | \begin{array}{cccc}
1 & 0 & 0 & 0 \\
0 & e^{z} & 0 & 0 \\
0 & 0 & e^{z} & 0 \\
0 & 0 & 0 & 1
\end{array} \right | \; ;
\eqno(2.2)
$$

\noindent eq. (2.1) takes the form
$$
\left [  \gamma^{0} {\partial \over \partial t} + \gamma^{1}
e^{z}  {\partial \over \partial x }  + \gamma^{2}  e^{z}
{\partial \over \partial y }  +  \gamma^{3}\left  (  {\partial
\over \partial z }-1\right  )  + im  \right ] \Psi  =0 \; .
\eqno(2.3)
$$

\noindent Note that the addition of $ -1 $ about the operator $
\partial_{z} $ can be removed by substituting  $\Psi = e^{z} \psi
$. The following three operators
 $i\partial_{t},\; i\partial_{x},\;i\partial_{y}$  commute with the wave operator:  solutions can be
  searched  in the form
  $$
\Psi ^{\epsilon, k_{1}, k_{2}}  = e^{-i \epsilon t} \; e^{i k_{1}x} \; e^{ik_{2}y} \; \left |
\begin{array}{r} f_{1}(z) \\ f_{2}(z) \\   f_{3}(z) \\
f_{4}(z)
\end{array} \right | \; .
\eqno(2.4)
$$

\noindent Using the Dirac matrices in spinor basis, from (2.3) we
find e equations for $ f_{i} (z) $
$$
-i\epsilon \; f_{3} \; -\; i k_{1} e^{z} \; f_{4} \; - \;  k_{2} e^{z} \; f_{4} \; - \; ( \; {d
\over d  z }-1 ) \; f_{3} + im \; f_{1} = 0 \; ,
$$
$$
-i\epsilon \; f_{4} \; -\;  i k_{1} e^{z} \; f_{3} \; + \; k_{2} e^{z} \; f_{3} \; + \; ( \; {d
\over d z }-1 ) \; f_{4} + im \; f_{2} = 0 \; ,
$$
$$
-i\epsilon \; f_{1} \; +\;  i k_{1} e^{z} \; f_{2} \; + \; k_{2} e^{z} \; f_{2} \; + \;  ( \; {d
\over d z }-1 ) \; f_{1} + im\; f_{3} = 0 \; ,
$$
$$
-i\epsilon \; f_{2} \; +\;  i k_{1} e^{z} \; f_{1} \; - \; k_{2}
e^{z} \; f_{1} \; - \; ( \; {d \over d z }-1 ) \; f_{2} + im \;
f_{4} = 0 \; . $$ $$
 \eqno(2.5)
$$

There is a generalized helicity operator which commutes with the operator of the wave equation:
$$
\Sigma = {1 \over 2}  \left (
e^{z}   \gamma^{2}\gamma^{3} {\partial \over \partial x } + e^{z}
 \gamma^{3}\gamma^{1}  {\partial \over \partial y }  +
 \gamma^{1}\gamma^{2} ( {\partial  \over \partial z } - 1)  \right ).
\eqno(2.6)
$$

Using the substitution (2.4) in   the eigenvalues equation  $ \Sigma \; \Psi = p \; \Psi $ we obtain
$$
 k_{1} e^{z} \; f_{2} \; - \;  ik_{2} e^{z} \; f_{2} \; - \;i ( \; {d
\over d z }-1 ) \; f_{1} =  p f_{1} \; ,
$$
$$
k_{1} e^{z} \; f_{1} \; +\;  ik_{2} e^{z} \; f_{1} \; + \;i ( \; {d
\over d z }-1 ) \; f_{2} =  p f_{2} \; ,
$$
$$
k_{1} e^{z} \; f_{4} \; - \;  ik_{2} e^{z} \; f_{4} \; - \;i ( \; {d
\over d z }-1 ) \; f_{3} =  p f_{3} \; ,
$$
$$
k_{1} e^{z} \; f_{3} \; + \;  ik_{2} e^{z} \; f_{3} \; + \;i ( \; {d
\over d z }-1 ) \; f_{4} =  p f_{4} \;.
 \eqno(2.7)
$$

\noindent From equations (2.7) and (2.5), considered together, it follows
 a linear homogeneous system with respect to $ f_ {i} $
$$
-i\epsilon\; f_{3} \; - \; ip f_{3} \; + \; im \; f_{1} =
0 \; ,
$$
$$
-i\epsilon\; f_{4} \; - \; ip \; f_{4} \; + \; im \; f_{2} =
0 \; ,
$$
$$
-i\epsilon\; f_{1} \; + \; ip \; f_{1} \; + \; im \; f_{3} =
0 \; ,
$$
$$
-i\epsilon\; f_{2} \; + \; ip \; f_{2} \; + \; im \; f_{4} =
0 \; . \eqno(2.8)
$$

\noindent We we find
two values for the $ p $ and the corresponding restrictions on the functions $ f_{i} $:
$$
p = \pm  \sqrt{\epsilon ^{2}  - m^{2}} \;  ,\qquad
f_{3} = {\epsilon - p \over m } \; f_{1} \; , \qquad  f_{4} =
{\epsilon- p \over m } \; f_{2} \; .
\eqno(2.9)
$$

Thus, we have three continuous quantum number $ \epsilon, \, k_{1}, \, k_{2} $ and one  discrete,
which  distinguishes the  values $p = \pm \sqrt{\epsilon ^{2}  - m^{2}}$.
In view of (2.9), from  four equations (2.5) we arrive at
two equations for $ f_{1}, \; f_{2} $
$$
( \; {d
\over d z }-1-i p  ) \; f_{1} +e^{z} (\;  i k_{1}   +   k_{2} ) \; f_{2} \;   = 0 \; ,
$$
$$
( \; {d
\over d z }-1 +ip ) \; f_{2}  -  e^{z} (\; i k_{1}   -  k_{2}  \;) f_{1} \;   = 0 \; .
 \eqno(2.10)
$$

\noindent
Note the symmetry of the equations with respect to change
$$
f_{1} \Longrightarrow f_{2} \;, \qquad  p \Longrightarrow - p \,.
\eqno(2.11)
$$

It is convenient to obtain  solutions of  similar  equations in the flat space
$$
( \; {d
\over d z }-i p  ) \; f_{1} + (\;  i k_{1}   +   k_{2} ) \; f_{2} \;   = 0 \; ,
$$
$$
( \; {d
\over d z } +ip ) \; f_{2}  -   (\; i k_{1}   -  k_{2}  \;) f_{1} \;   = 0 \; ,
 \eqno(2.12)
$$

\noindent so that
$$
f_{2} = - {1 \over ik_{1} + k_{2}} ( {d \over dz} - i p ) f_{1} \; ,
$$
$$
\left ( {d^{2} \over dz^{2}} + \epsilon^{2} - m^{2} - k_{1}^{2} - k_{2}^{2} \right ) f_{1} =0 \,.
\eqno(2.13)
$$

\noindent
Here, there exist two independent solutions (let $k_{3} = +
\sqrt{\epsilon^{2} - m^{2} - k_{1}^{2} - k_{2}^{2}}$)
$$
f^{(1)} _{1} = e^{+ik_{3} z} \;, \qquad
f_{2} ^{(1)} = - {( + i k_{3}  - i p )  \over ik_{1} + k_{2}} \;  e^{+ik_{3} z}  \; ,
$$
$$
f^{(2)} _{1} = e^{-ik_{3} z} \;,
\qquad
f_{2}^{(2)} = - {( -i k_{3}  - i p )  \over ik_{1} + k_{2}}  \; e^{-ik_{3} z}  \; .
\eqno(2.14)
$$

\noindent
The sign before $ k_{3} $ determines the direction of  the wave propagation,
the sign of $ p $ defines the state of polarization.
Generalized analogue of these solution is  to be investigated in the  hyperbolic space $ H_{3} $.

\subsection*{3  A special case of the waves along the  $z$-axis}

\hspace{7mm}There exists  a special  case when $k_{1}=0, \,k_{2}=0$:
$$
\Psi^{\epsilon,0,0}(t,z)   = e^{-i \epsilon t} \;  \left |
\begin{array}{r} f_{1}(z) \\ f_{2}(z) \\   f_{3}(z) \\
f_{4}(z)
\end{array} \right |  . \eqno(3.1)
$$

\noindent The equations change substantially  (see  (2.5))
$$
-i\epsilon \; f_{3} \;  - \; ( \; {d
\over d z }-1 ) \; f_{3} + im \; f_{1} = 0 \; ,
$$
$$
-i\epsilon \; f_{4} \;
 + \; ( \; {d
\over d z }-1 ) \; f_{4} + im \; f_{2} = 0 \; ,
$$
$$
-i\epsilon \; f_{1} \; + \;  ( \; {d
\over d z }-1 ) \; f_{1} + im\; f_{3} = 0 \; ,
$$
$$
-i\epsilon \; f_{2} \;  - \; ( \; {d
\over d z }-1 ) \; f_{2} + im \; f_{4} = 0 \; ,
 \eqno(3.2)
$$

\noindent
  diagonalization of the operator $ \Sigma $ gives (see (2.7))
$$
( \; {d
\over d z }-1 ) \; f_{1} = i p f_{1} \; , \qquad
( \; {d
\over d z }-1 ) \; f_{2} =  -i p f_{2} \; ,
$$
$$
(  {d
\over d z }-1 ) \; f_{3} =  i p f_{3} \; , \qquad
 (  {d
\over d z }-1 ) \; f_{4} =  -i p f_{4} \;.
 \eqno(3.3)
$$

\noindent Considering  equations (3.2) and (3.3) together, we arrive at
the linear system
$$
  ( \epsilon  + p ) \; f_{3}  - m \; f_{1} = 0 \; , \qquad
 (\epsilon  + p ) \; f_{4}   -  m \; f_{2} = 0 \; ,
$$
$$
 (\epsilon - p ) \; f_{1}  - m\; f_{3} = 0 \; ,
\qquad
 (\epsilon - p ) \; f_{2}    - m \; f_{4} = 0 \; .
 \eqno(3.4)
$$

\noindent
From this it follows
$$
(\epsilon^{2} - p^{2})^{2} - m^{4} =0 \qquad \Longrightarrow \qquad
 p = \pm \sqrt{\epsilon^{2} - m^{2}} \; ,
$$
$$
f_{4} = {\epsilon - p \over m} f_{2} = {m \over  \epsilon + p } f_{2} \; , \qquad
f_{3} = {\epsilon - p \over m} f_{1} = {m \over  \epsilon + p } f_{1} \; .
 \eqno(3.5)
$$

\noindent Equation (3.2)  reduce to
$$
( \; {\partial
\over \partial z }-1-i p  ) \; f_{1}     = 0  \qquad \Longrightarrow \qquad
f_{1} = C_{1} \; e^{z} e^{+i p z} \; ,
$$
$$
( \; {\partial
\over \partial z } -1 +ip ) \; f_{2}     = 0  \qquad \Longrightarrow \qquad
 f_{2} = C_{2} \; e^{z} e^{-i p z} \; .
 \eqno(3.6)
$$

 Solutions more simple  to  interpret are
$$
C_{1} = 1\;, \qquad  C_{2} = 0\;,\qquad
\Psi^{\epsilon 00 p}(t,z) =
\left |
\begin{array}{c}
1 \\
0\\
{\epsilon - p\over m}
\\
0
\end{array} \right | e^{z}e^{+ip z} ;
\eqno(3.7a)
$$
$$
C_{1} = 0\;, \qquad  C_{2} = 1\;,\qquad
\Psi^{\epsilon 00 p}(t,z) =
\left |
\begin{array}{c}
0
 \\
 1\\
0
  \\
{\epsilon - p\over m}
\end{array} \right | e^{z}e^{-ip z} .
\eqno(3.7b)
$$

Obviously, the factor $ e ^ {z} $  in the solutions  will be compensated
when considering any bilinear structure of the wave functions (with their subsequent multiplication  by
$\sqrt{-g}\; dxdydz$).

\subsection*{4  Construction of solutions  in the general case}

 \hspace{5mm}
Let us turn to (2.10)
and introduce a new variable
$Z$:
$$
\sqrt{k_{1}^{2} + k_{2}^{2}}\;  e^{z} = Z
\;, \qquad Z \in (0, + \infty )\;,
$$
$$
\left (   Z {d \over d Z}  -1-i p \right  )  f_{1} +
 Z \sqrt{{ k_{2} + i k_{1} \over k_{2} - i k_{1} }}
  \; f_{2} \;   = 0 \; ,
$$
$$
\left (  Z {d \over d Z} -1 +ip \right )  f_{2}  +  Z  \sqrt{{ k_{2} - i k_{1} \over
k_{2} + i k_{1} }}
  \; f_{1} \;   = 0 \; .
 $$

\noindent
It is convenient to define a parameter
$$
e^{ia} = \sqrt{{ k_{2} + i k_{1} \over k_{2} - i k_{1} }}\,,
$$

\noindent
then
$$
\left (   Z {d \over d Z}  -1-i p  \right  )  f_{1} +
 Z  e^{+ia}
   f_{2}    = 0 \; ,
$$
$$
\left (  Z {d \over d Z} -1 +ip  \right )  f_{2}  +   Z e^{-ia}
   f_{1}    = 0 \; .\eqno(4.1a)
 $$

\noindent
Note that the factor $ e ^ {i \alpha} $ can be removed by changing the notation
for  $f_{2}$
$$e^{+ia}    f_{2} \;\; \Longrightarrow \;\; f_{2} \; ;
$$

\noindent
then the  system   takes the form
$$
\left (   Z {d \over d Z}  -1-i p  \right  )  f_{1} +
 Z
   f_{2}    = 0 \; ,
$$
$$
\left (  Z {d \over d Z} -1 +ip  \right )  f_{2}  +   Z
   f_{1}    = 0 \; .\eqno(4.1b)
 $$

From  $ (4.1a) $ we get  two second order differential equations for $f_{1}$ and $f_{2}$:
$$
Z\,{\,d^{2}f_{1}\over dZ^{2}}-2\,{\,d\,f_{1}\over dZ}+\left({p^{2}+ip+2\over Z}-Z\right)f_{1}=0\,,
\eqno(4.2a)
$$
$$
Z\,{\,d^{2}f_{2}\over dZ^{2}}-2\,{\,d\,f_{2}\over dZ}+\left({p^{2}-ip+2\over Z}-Z\right)f_{2}=0\,.
\eqno(4.2b)
$$

\noindent Note the symmetry between the equations: they  are transformed into each other
when changing $p \longrightarrow- p$.
Furthermore, it should be noted that unlike the case of flat space (see (2.13)),
 here second-order equations for the functions $ f_ {1} $ and $ f_ {2} $
  depend explicitly on the  first degree of
 $p$,  that is depends on the  state  of polarization of  spinor waves.

Considering  eq. $ (4.2a) $, let us use a substitution $f_{1}(Z)=Z^{A}e^{BZ}F_{1}(Z)$:
$$
Z\,{\,d^{2}F_{1}\over dZ^{2}}+(2A-2+2BZ)\,{\,d\,F_{1}\over dZ}+
$$
$$
+\left[(B^{2}-1)Z+2B\,(A-1)+{(A-ip-1)\,(A+ip-2)\over Z}\right]F_{1}=0\,.\eqno(4.3)
$$

\noindent
At  $ A $ and $ B $ chosen according
$$
A= + ip+1\,,\;\;-ip+2\,,\qquad B=\pm 1\,,\eqno(4.4)
$$

\noindent (4.3)  becomes simpler
$$
Z\,{\,d^{2}F_{1}\over dZ^{2}}+(2A-2+2BZ)\,{\,d\,F_{1}\over dZ}+2B\,(A-1)F_{1}=0\,.\eqno(4.5)
$$

\noindent
The resulting equation with  another change of $ Z = y / 2 $ will transform into
$$
y\,{\,d^{2}F_{1}\over dy^{2}}+(2A-2+By)\,{\,d\,F_{1}\over dy} + B\,(A-1)F_{1}=0\,.\eqno(4.6a)
$$

\noindent
When $B=-1$, it coincides with  an equation for the confluent
hypergeometric function with parameters (for definiteness let it be    $A= +ip+1 $)
$$
y\,{\,d^{2}\Phi \over dy^{2}}+(c -y) {\,d\,\Phi \over dy}- a \Phi =0\, , \qquad
a  = + i p \,,\qquad c =2 a = + 2ip  \,.
\eqno(4.6b)
$$

Two linearly independent solutions are \cite{Beytmen-1973}
$$
 F _{1}^{(1)}(y) = \Phi (a,c,y)  \; ,
 $$
 $$
F _{1}^{(2)} (y)  = y^{1-c} \Phi (a-c+1,2-c,y)  \;.
\eqno(4.6c)
$$


Consider the equation $(4.2b)$. Using the above-noted symmetry, we obtain
$$
f_{2} = y^{a'+1} e^{-y/2} F_{2} (y) \; , \qquad a' = -i p \; , \; c' = 2a' = -2i p \; ,
$$
$$
 F _{2}^{(1)} = \Phi (a',c',y)  \; ,
 $$
 $$
F _{2}^{(2)}  = y^{1-c'} \Phi (a'-c'+1,2-c',y) \; .
\eqno(4.7)
$$

It is convenient to employ one  independent parameter $a$:
$$
 f_{1} = y^{a+1} e^{-y/2} F_{1} (y)\;  ,
 $$
 $$
 F _{1}^{(1)}(y) = \Phi (a,2a,y)  \; ,
 $$
 $$
F _{1}^{(2)} (y)  = y^{1-2a} \Phi (1-a,2-2a,y)  \; ;
\eqno(4.8a)
$$

$$
 f_{2} = y^{-a+1} e^{-y/2} F_{2} (y) \;  ,
 $$
 $$
 F _{2}^{(1)} (y) = \Phi (-a,-2a,y)  \; ,
 $$
 $$
F _{2}^{(2)} (y)  = y^{1+2a} \Phi (1+a,2+2a,y) \; .
\eqno(4.8b)
$$

The functions $ f_ {1}, \, f_ {2} $  (note that before now  we did not  find   possible numerical factors at them)
must be  related by the  first-order operators (see  (4.1))
$$
\left (   y {d \over d y}  -1- a  \right )  f_{1} - {y \over 2}  e^{+ia}
   f_{2}    = 0 \; ,
 $$
 $$
 \left (  y {d \over d y} -1 + a  \right )  f_{2}  - {y \over 2}  e^{-ia}
   f_{1}    = 0 \;;
 $$

\noindent  these  relationships can be translated to
the functions $ F_ {1}, \, F_ {2} $, which results in
$$
{dF_{1}\over dy}-{1\over
2}\,\left(F_{1}+y^{-2a}\,e^{+ia}F_{2}\right)=0\,,\eqno(4.9a)
$$
$$
{dF_{2}\over dy}-{1\over
2}\,\left(F_{2}+y^{+2a}\,e^{-ia}F_{1}\right)=0\,.\eqno(4.9b)
$$

\noindent
These equations relate  functions in the  following pairs
$$
F_{1}^{(1)}(y) --- F_{2}^{(2)}(y) \; , \qquad F_{1}^{(2)}(y) --- F_{2}^{(1)}(y)\;.
$$

For each pair one should  find the relative coefficient of the two functions.

 Let us  substitute  expressions for
$ F_ {1} ^ {(1)} (y) $ and $ F_ {2} ^ {(2)} (y) $ with some numerical coefficients
$$
 F _{1}^{(1)}(y) = r_{1}^{(1)}\,\Phi (a,2a,y)\,,
 \qquad
 F _{2}^{(2)} (y)  = r_{2}^{(2)}\,y^{1+2a} \Phi (a+1,2+2a,y)
$$

\noindent in eq.  $(4.9a)$. Performing the necessary differentiation
$$
r_{1}^{(1)} \left [ \Phi\,(a+1, 2a+1, y)- \Phi\,(a, 2a,
y) \right ] -
 r_{2}^{(2)}\,e^{+ia}\,y\,\Phi\,(a+1, 2a+2, y)=0,
$$

\noindent and transformation using the relations for contiguous functions, we find
the relative factor
$$
r_{1}^{(1)}=2\,e^{+ia}\,(2a+1 )\,r_{2}^{(2)}\,.
\eqno(4.10)
$$

\noindent
The same result can be obtained by substituting the expression for
$ F_ {1} ^ {(1)} (y) $ and $ F_ {2} ^ {(2)} (y) $ in the equation $ (4.9b) $.
The resulting ratio can also be obtained by using the expansions of solutions near $ y = 0 $
$$
F_{1}^{(1)} = r_{1}^{(1)} (1 + {1 \over 2}  y + { (a+1) \over 2 (2a+1)}  y^{2} + ...,
\qquad F _{2}^{(2)} (y)  = r_{2}^{(2)}\,y^{1+2a} \; ;
$$
the equation
$$
{dF_{1}^{(1)} \over dy} - {1\over 2} F_{1}^{(1)} - {1\over 2} y^{-2a}\,e^{+ia} F_{2}^{(2)} =0
$$

\noindent
gives
$$
r_{1}^{(1)} \left ( {1 \over 2} + {a+1 \over 2a+1} y -{1 \over 2} - {1\over 2} y \right ) -
e^{+i\alpha} r_{2}^{(2)} y = 0
\; \;\; \Longrightarrow
$$
$$
r_{1}^{(1)} {1 \over 2(2a+1)} - e^{+i\alpha} r_{2}^{(2)} = 0\; .
$$

Now let us substitute  expressions for $ F_ {1} ^ {(2)} (y) $ and
$ F_ {2} ^ {(1)} (y) $ of $ (4.8) $
$$
F _{1}^{(2)} (y)  =  r _{1}^{(2)} y^{1-2a} \Phi (1-a,2-2a,y)\,,
$$
$$
F _{2}^{(1)} (y) =  r _{2}^{(1)} (y) \Phi (-a,-2a,y)
\eqno(4.11a)
$$

\noindent in the equation  $(4.9b)$. Performing the differentiation
$$
r_{1}^{(2)} \left [ \Phi\,(1-a, 1-2a, y)- \Phi\,(-a, -2a,
y) \right ] -
$$
$$
-
r_{2}^{(1)}\,e^{-ia}\,y\,\Phi\,(1-a, 2-2a, y)=0
$$

\noindent and the corresponding transformations, we arrive at
$$
r_{1}^{(2)}=2\,e^{-ia}\,(1-2a)\,r_{2}^{(1)}\,.
\eqno(4.11b)
$$

Thus, the construction of two linearly independent solutions of (4.1):
$$
f_{1} =  r_{1}^{(1)} \;  e^{-y/2} y^{a+1}  \,\Phi (a,2a,y) \; ,\qquad
$$
$$
f_{2} = r_{2}^{(2)}\;  e^{-y/2}   y^{2+a} \Phi (a+1,2+2a,y)\; ,
$$
$$
r_{1}^{(1)}=2\,e^{+ia}\,(1+ 2a )\,r_{2}^{(2)}  \;; \qquad \qquad
\eqno(4.12a)
$$

$$
f_{2} = r _{2}^{(1)}  \;  e^{-y/2}  y^{-a+1}   \Phi (-a,-2a,y)  \; , \;\;\;\;
$$
$$
f_{1} = r _{1}^{(2)}  \;  e^{-y/2}   y^{2-a} \Phi (1-a,2-2a,y)\; ,
$$
$$
r_{1}^{(2)}=2\,e^{-ia}\,(1-2a)\,r_{2}^{(1)} \;. \qquad \qquad
\eqno(4.12b)
$$

We simplify the formulas by setting
$
r_{2}^{(1)}= 1\; , \; r_{2}^{(2)}= 1;
$
the result is
$$
I \qquad f_{1} = M_{+}   \;  e^{-y/2} y^{1+a}  \,\Phi (a,2a,y) \; ,
$$
$$
\qquad  \;\;\;\; f_{2} =  e^{-y/2}   y^{2+a} \Phi (a+1,2+2a,y) \; ,
$$
$$
M _{+} =  \left [  2\,e^{+ia}\,(1+ 2a ) \right ] \; ;
\eqno(4.13a)
$$

$$
II \qquad  \qquad\qquad f_{1} = M_{-}    \;  e^{-y/2}   y^{2-a} \Phi (1-a,2-2a,y) \; ,
$$
$$
f_{2} =   e^{-y/2}  y^{1-a}   \Phi (-a,-2a,y)
 \; ,
$$
$$
M_{-} = \left [ 2\,e^{-ia}\,(1-2a) \right ] \; .
\eqno(4.13b)
$$

Remind that $a = i p = \pm i \sqrt{\epsilon^{2}-m^{2}}$; the sign of $ p $
is associated with the polarization state of the spinor waves;
types $ I $ and $ II $ are supposed to be associated with the directions of
wave propagation: to the left or to the right.

Let us consider  asymptotic properties of the solutions.
The coordinate $ y = 2 \sqrt {k_ {1} ^ {2} + k_ {2} ^ {2}} \;
e ^ {z} $ tends to  (see (1.2))
$$
q \longrightarrow  -1 \; , \qquad y= 2 \sqrt{k_{1}^{2} + k_{2}^{2}} \; e^{z}
  \;\; \longrightarrow \;\; 0 \; ,
$$
$$
q \longrightarrow  +1 \; ,  \qquad y = 2 \sqrt{k_{1}^{2} + k_{2}^{2}} \; e^{z}
  \;\; \longrightarrow \;\;  + \infty .
\eqno(4.13c)
$$

\noindent The behavior of solutions in  boundary point $  y \rightarrow 0\; (z \rightarrow - \infty )$:

\vspace{2mm}

$I$
$$
f_{1} = M_{+}   \;   y^{1+a}  = M_{+}   \;  \left  (2 \sqrt{k_{1}^{2} +k_{2}^{2}} \; e^{z} \right )^{1+i p }  \; ,
$$
$$
f_{2} =    y^{2+a}   = \left  (2 \sqrt{k_{1}^{2} +k_{2}^{2}} \; e^{z} \right )^{2+i p } \; ;
\eqno(4.14a)
$$

$II$
$$
f_{1} = M_{-}    \;   y^{2-a} = \left  (2 \sqrt{k_{1}^{2} +k_{2}^{2}} \; e^{z} \right )^{2-i p} \; ,
$$
$$
f_{2} =    y^{1-a}   = \left  (2 \sqrt{k_{1}^{2} +k_{2}^{2}} \; e^{z} \right )^{1-i p } \; . \eqno(4.14b)
$$

For large values of $ y $ one should  use the asymptotic formula
 \cite{Beytmen-1973}
$$
y \rightarrow + \infty \;, \qquad \Phi (A,C,y) = {\Gamma (C) \over \Gamma (A)}e^{y} y^{A-C} \; .
$$

\noindent So we get (for \underline{$ z \rightarrow +\infty, \qquad y \rightarrow + \infty$})
$$
I\qquad \qquad f_{1} =  M_{+}   \;  e^{y/2} y  \,{\Gamma (2a) \over \Gamma (a)}\, , \qquad
f_{2} =  e^{y/2} y  \,{\Gamma (2+2a) \over \Gamma (a+1)}\, ,
$$
$$
\eqno(4.15a)
$$
$$
II \qquad \qquad
f_{1} =  M_{-}\,e^{y/2} y  \,{\Gamma (2-2a) \over \Gamma (1-a)} \, ,
\qquad
f_{2} =   e^{y/2} y  \,{\Gamma (-2a) \over \Gamma (-a)}\, .
$$
$$
\eqno(4.15b)
$$

To conclude this section we consider the limiting process in the constructed solutions $ (4.13a), \,
 (4.13b) $ to the case of  the flat space.
This will allow a better understanding of  the  obtained  results in  the Lobachevsky space.

To this end,  we first need to go to the usual dimensional quantities:
$$
 z = {z_{3} \over R} \; ,
\qquad m = {Mc R \over \hbar} \;, \qquad \epsilon =  {E R \over  c \hbar }\;,
$$

$$
p = + \sqrt{\epsilon^{2} - m^{2}} = +
R \; \sqrt{ E^{2} /  c^{2} \hbar^{2}  - M^{2} c^{2} /  \hbar^{2}}  =  R p_{0}  \;,
$$

$$
k_{1} = {P_{1} R \over c \hbar }\;, \qquad
k_{2} = {P_{2} R \over c \hbar }\;, \qquad \sqrt{k_{1}^{2} + k_{2}^{2}} =
 R{ \sqrt{P_{1}^{2} + P_{2}^{2}} \over c \hbar } = R K_{\bot}\;,
$$

$$
a = i p =   i R p_{0},
\qquad
c =2a = i 2 R p_{0}\,,
$$

$$
y = 2 \sqrt{k_{1}^{2} + k_{2}^{2}}  e^{z} = 2  R  K_{\bot}
 ( 1 + {x_{3} \over  R} + ...) \qquad \longrightarrow \qquad    2 R K_{\bot}\,.
$$
$$
\eqno(4.16)
$$

Let us consider the solutions $(4.13a)$
$$
I \qquad f_{1} = M_{+}   \;  e^{-y/2} y^{1+a}  \,\Phi (a,2a,y) \; ,
$$
$$
\qquad  \;\;\;\; f_{2} =  e^{-y/2}   y^{2+a} \Phi (a+1,2+2a,y) \; ,
$$
$$
M _{+} =  \left [  2\,e^{+ia}\,(1+ 2a ) \right ] \; ;
$$

\noindent taking into account
$$
{a \over c} y = {1 \over 2 } y \qquad \Longrightarrow \qquad R K_{\bot}\; ,
$$
$$
{1 \over 2!} {a (a+1) \over c(c+1) } y^{2}  = {1 \over 2!} {1/2 (1/2+1/c) \over (1+1/c ) } y^{2}
 \qquad \Longrightarrow \qquad  {1 \over 2!} \; (\;  R K_{\bot} \; ) ^{2}  \; ,
$$
$$
{1 \over 3!} {a (a+1)(a+2) \over c(c+1)(c+2)  } y^{2}  = {1 \over 2!} {1/2 (1/2+1/c)(1/2 +2/c)
 \over (1+1/c )(1 +2/c)  } y^{2}
 \qquad \Longrightarrow
 $$
 $$
 \qquad  {1 \over 3!} \; (\;  R K_{\bot} \; ) ^{3}  \; ... ,
\eqno(4.17)
$$

\noindent we get
$$
e^{-y/2} \;\; \Longrightarrow \;\; e^{-RK_{\bot}} \;, \qquad
\Phi (a,2a,y) \qquad \Longrightarrow \qquad  e^{RK_{\bot}} \,,
$$
$$
e^{-y/2} \;\; \Longrightarrow \;\; e^{-RK_{\bot}} \;, \qquad
\Phi (a+1,2a+2,y) \qquad \Longrightarrow \qquad  e^{RK_{\bot}} \,,
$$

\noindent and further
$$
I \qquad f_{1}  \qquad \Longrightarrow \qquad   M_{+}   \;   (2 R K_{\bot} e^{z} )^{1+iR p_{0}}
 \,      \sim e^{ix_{3} p_{0}} \; ,
$$
$$
\qquad  \;\;\;\; f_{2} \qquad \Longrightarrow \qquad
(2 R K_{\bot} e^{z} )^{2+iR p_{0}}  \sim e^{ix_{3} p_{0}}  \,.
\eqno(4.18)
$$

\noindent
Similarly, we find
$$
II \qquad f_{2} =   e^{-y/2}  y^{1-a}   \Phi (-a,-2a,y) \sim  e^{-ix_{3} p_{0}}  \; ,
$$
$$
\qquad\qquad f_{1} = M_{-}    \;  e^{-y/2}   y^{2-a} \Phi (1-a,2-2a,y) \sim e^{-ix_{3} p_{0}}  \; .
$$
$$
\eqno(4.19)
$$

We may conclude that solutions of the type I (in curved model  $H_{3}$) provide us with extension for the flat waves
in Minkowski space of the type $e^{+ikz}$; whereas solutions of the type II
represent extension for the flat waves
in Minkowski space of the type $e^{-ikz}$.

\subsection*{5 The case of the Weyl neutrino}

\hspace{5mm}
Let us  restrict the above  analysis to the case of  2-component Weyl neutrino.
It is enough to specify  the Dirac equation in the spinor basis (see \cite{Red'kov-book-2009})
$$
i\; \sigma ^{\alpha }(x) \; [\; \partial _{\alpha }
 \; + \; \Sigma _{\alpha }(x)\; ] \; \xi (x) = \; m \; \eta (x)  \;   ,
$$
$$
i\; \bar{\sigma }^{\alpha }(x) \; [ \;\partial _{\alpha } \; + \;
\bar{\Sigma }_{\alpha }(x) \;] \; \eta (x) = \; m \; \xi (x)
$$

\noindent  and impose  relation $m = 0$, thus we will obtain the equations of the Weyl
   neutrino (wave function $ \eta (x) $) and
antineutrinos (with the wave function  $\xi(x)$).

Substitution for  $ \eta $ is
$$
\eta ^{\epsilon, k_{1}, k_{2}}  = e^{-i \epsilon t} \; e^{i k_{1}x} \; e^{ik_{2}y} \; \left |
\begin{array}{r}   f_{3}(z) \\
f_{4}(z)
\end{array} \right | \; .
\eqno(5.1)
$$

\noindent
Note that neutrino wave function   has only three quantum numbers $\epsilon, \,k_{1},\, k_{2}$.
The system of equations after separation of variables has the form
$$
-i\epsilon \; f_{3}  - i k_{1} e^{z}  f_{4}  -   k_{2} e^{z}  f_{4}  -  (  {\partial
\over \partial z }-1 ) \; f_{3}  = 0 \; ,
$$
$$
-i\epsilon \; f_{4}  -  i k_{1} e^{z}  f_{3}  +  k_{2} e^{z}  f_{3}  +  (  {\partial
\over \partial z }-1 ) \; f_{4} = 0 \;
 \eqno(5.2)
$$

\noindent
or (replacing $ f_ {3} $ on $ h_ {1} $ and $ f_ {4} $ on $ h_ {2}$)
$$
( {d \over d z} -1 + i \epsilon )\,h_{1} + e^{z} (ik_{1} + k_{2})\, h_{2} = 0 \; ,
$$
$$
( {d \over d z} -1 - i \epsilon )\,h_{2} -  e^{z} (ik_{1} -k_{2})\, h_{1} = 0 \; .\eqno(5.3)
$$

\noindent
One can use the solution obtained above (for the system (2.10)), replacing everywhere  $ p $ on $ - \epsilon$.
Thus, we get two linearly independent solutions of the system (5.3)
$$
h_{1} =  r_{1}^{(1)} \;  e^{-y/2} y^{a+1}  \,\Phi (a,2a,y) \; ,\qquad
$$
$$
h_{2} = r_{2}^{(2)}\;  e^{-y/2}   y^{2+a} \Phi (a+1,2+2a,y)\; ,
$$
$$
r_{1}^{(1)}=2\,e^{+ia}\,(1+ 2a )\,r_{2}^{(2)}  \;; \qquad \qquad
\eqno(5.4a)
$$

$$
h_{2} = r _{2}^{(1)}  \;  e^{-y/2}  y^{-a+1}   \Phi (-a,-2a,y)  \; , \;\;\;\;
$$
$$
h_{1} = r _{1}^{(2)}  \;  e^{-y/2}   y^{2-a} \Phi (1-a,2-2a,y)\; ,
$$
$$
r_{1}^{(2)}=2\,e^{-ia}\,(1-2a)\,r_{2}^{(1)} \,, \qquad \qquad
\eqno(5.4b)
$$

\noindent where
$$
a  =- i \epsilon \,,\qquad c =2 a = -2i\epsilon \,.
\eqno(5.4c)
$$

\subsection*{6 On the non-relativistic Pauli approximation}

 \hspace{5mm}
We will carry out a procedure of the  non-relativistic approximation
   directly in the separated  equations for relativistic Dirac case.
To this end,  we turn to equations (2.5)
$$
-i\epsilon \; f_{2} \; +\;  i k_{1} e^{z} \; f_{1} \; - \; k_{2} e^{z} \; f_{1} \; - \; ( \; {d
\over d z }-1 ) \; f_{2} + im \; f_{4} = 0 \; ,
$$
$$
-i\epsilon \; f_{4} \; -\;  i k_{1} e^{z} \; f_{3} \; + \; k_{2} e^{z} \; f_{3} \; + \; ( \; {d
\over d z }-1 ) \; f_{4} + im \; f_{2} = 0 \; ,
$$
$$
 -i\epsilon \; f_{3} \; -\; i k_{1} e^{z} \; f_{4} \; - \;  k_{2} e^{z} \; f_{4} \; - \; ( \; {d
\over d  z }-1 ) \; f_{3} + im \; f_{1} = 0 \; ,
$$
$$
-i\epsilon \; f_{1} \; +\;  i k_{1} e^{z} \; f_{2} \; + \; k_{2} e^{z} \; f_{2} \; + \;  ( \; {d
\over d z }-1 ) \; f_{1} + im\; f_{3} = 0\; .
$$
$$
 \eqno(6.1)
$$

\noindent We introduce new functions
$$
{f_{1} + f_{3} \over 2  } = f \;, \qquad {f_{1} - f_{3} \over 2 i }
= g  \;,
$$
$$
{f_{2} + f_{4} \over 2 } = F \;, \qquad {f_{2} - f_{4} \over 2  i}
= G  \;.
\eqno(6.2)
$$

\noindent From (6.1), one
obtains  equations for  $f , \; F , \;  g , \; G :$
$$
  ( {d \over d z }  -  1  ) \; G
 -e^{z}\,(ik_{1}-k_{2})\;g+ (   \epsilon  -m  )\,F =0\, ,
$$
$$
  ( {d \over d z }  -  1  ) \; F
 -e^{z}\,(ik_{1}-k_{2})\;f- (   \epsilon  +m  )\,G =0\, ,
$$
$$
  ( {d \over d z }  -  1  ) \; g
+e^{z}\,(ik_{1}+k_{2})\;G- (   \epsilon  -m  )\,f =0\, ,
$$
 $$
  ( {d \over d z }  -  1  ) \; f
 +e^{z}\,(ik_{1}+k_{2})\;F+ (   \epsilon  +m  )\,g =0\,.
\eqno(6.3)
 $$

 Now we  should make a formal change
 $ \epsilon  \; \Longrightarrow \;  m + E $
 (this is equivalent to separating the rest energy  by a factor $e^{-imt}$); as a result we get

 $$
  ( {d \over d z }  -  1  ) \; G
 -e^{z}\,(ik_{1}-k_{2})\;g+ E\,F =0\, ,
$$
$$
  ( {d \over d z }  -  1  ) \; F
 -e^{z}\,(ik_{1}-k_{2})\;f- (   E  +2m  )\,G =0\, ,
$$
$$
  ( {d \over d z }  -  1  ) \; g
+e^{z}\,(ik_{1}+k_{2})\;G- E\,f =0\, ,
$$
 $$
  ( {d \over d z }  -  1  ) \; f
 +e^{z}\,(ik_{1}+k_{2})\;F+ (  E  +2m  )\,g =0\,.
\eqno(6.4)
 $$

The condition for the applicability of the nonrelativistic approximation is
the following relation
$
E+ 2m \approx  2 m ;
$ which results in

$$
  ( {d \over d z }  -  1  ) \; G
 -e^{z}\,(ik_{1}-k_{2})\;g+ E\,F =0\, ,
$$
$$
G   = {1\over 2m}\,\left[( {d \over d z }  -  1  ) \; F
 -e^{z}\,(ik_{1}-k_{2})\;f\right] ,
$$
$$
  ( {d \over dz }  -  1  ) \; g
+e^{z}\,(ik_{1}+k_{2})\;G- E\,f =0\, ,
$$
 $$
g =-{1\over 2m}\,\left[( {d \over d z }  -  1  ) \; f
 +e^{z}\,(ik_{1}+k_{2})\;F\right].
\eqno(6.5)
 $$

Excluding  two  small components of $ g, \, G $, we arrive at two equations for the big components
$f$  and $F$:

$$
\left[{d^{2}\over dz^{2}}-2\,{d\over dz}+1-e^{2z}\,(k_{1}^{2}+k_{2}^{2})+2\,m\,E\right]F-e^{z}(i\,k_{1}-k_{2})\;f=0\,,
$$
$$
\left[{d^{2}\over dz^{2}}-2\,{d\over dz}+1-e^{2z}\,(k_{1}^{2}+k_{2}^{2})+2\,m\,E\right]f+e^{z}(i\,k_{1}+k_{2})\;F=0 \,.
$$
$$
\eqno(6.6)
$$

We remind  that (see (2.9))

$$
f = {f_{1} (z) + f_{3} (z) \over 2} = {1 +A \over 2} \,f_{1} (z)  \;,
$$
$$
F = {f_{2} (z) + f_{4} (z) \over 2} = {1+A \over 2} \,f_{2} (z)  \;,
$$

$$
A={\epsilon\pm p\over m}\,, \qquad p = \pm \sqrt{\epsilon^{2} - m^{2}} =
$$
$$
=  \pm
 \sqrt{(E+m)^{2} - m^{2}} \approx  \pm \sqrt{2mE} \; ;
\eqno(6.7a)
$$

\noindent sign $ \pm $ correspond to two different polarizations of the non-relativistic electron. Therefore, equation (6.6) can be represented as follows (remember that now the parameter $ p $ defined by the nonrelativistic expression
$ (6.7a)$):
$$
\left[{d^{2}\over dz^{2}}-2\,{d\over dz}+1-e^{2z}\,(k_{1}^{2}+k_{2}^{2})+2\,m\,E\right]f_{2}-e^{z}(i\,k_{1}-k_{2})\;f_{1}=0\,,
$$
$$
\left[{d^{2}\over dz^{2}}-2\,{d\over dz}+1-e^{2z}\,(k_{1}^{2}+k_{2}^{2})+2\,m\,E\right]f_{1}+e^{z}(i\,k_{1}+k_{2})\;f_{2}=0 \,.
$$
$$
\eqno(6.7b)
$$

\noindent These equations can be written as (for definiteness, let $ p = + \sqrt {2ME}$)
$$
\left[( {d
\over d z }-1-i p)({d
\over d z }-1+i p)-e^{2z}\,(k_{1}^{2}+k_{2}^{2})\right]f_{2}-e^{z}(i\,k_{1}-k_{2})\;f_{1}=0\,,
$$
$$
\left[( {d
\over d z }-1-i p)({d
\over d z }-1+i p)-e^{2z}\,(k_{1}^{2}+k_{2}^{2})\right]f_{1}+e^{z}(i\,k_{1}+k_{2})\;f_{2}=0 \,.
$$
$$
\eqno(6.8)
$$

We take into account (2.10)
$$
( \; {d
\over d z }-1-i p  ) \; f_{1} +e^{z} (\;  i k_{1}   +   k_{2} ) \; f_{2} \;   = 0 \; ,
$$
$$
( \; {d
\over d z }-1 +ip ) \; f_{2}  -  e^{z} (\; i k_{1}   -  k_{2}  \;) f_{1} \;   = 0 \; .
$$

\noindent Thus we arrive at equations
$$
({d
\over d z }-1 -ip)\,(e^{z}\,(ik_{1}-k_{2})\,f_{1})-
$$
$$
-e^{z}(ik_{1}-k_{2})\,({d
\over d z }-1 -ip)\,f_{1}-e^{z}\,(ik_{1}-k_{2})\,f_{1}= 0\,,
$$
$$
({d
\over d z }-1 +ip)\,(-e^{z}\,(ik_{1}+k_{2})\,f_{2})+
$$
$$
+ e^{z}(ik_{1}+k_{2})\,({d
\over d z }-1 +ip)\,f_{2}+e^{z}\,(ik_{1}+k_{2})\,f_{2}= 0 \,.
$$
$$
\eqno(6.9)
$$

\noindent
It is easy to see that these two identities of the form $ 0 = 0$.

Thus, the solutions found above for the relativistic Dirac equation
in  the approximation $p \approx \sqrt {2mE}$ are solutions of two-component Pauli equation.

\subsection*{7 On representation of solutions in terms of Bessel functions}

\hspace{5mm}Returning to the basic system (4.1) in the form
$$
\left (   Z {d \over d Z}  -1-i p  \right  )  \sqrt{k_{2} - i k_{1}} f_{1} +
 Z  \sqrt{k_{2} + i k_{1}}
   f_{2}    = 0 \; ,
$$
$$
\left (  Z {d \over d Z} -1 +ip  \right )  \sqrt{k_{2} + i k_{1}} f_{2}  +   Z \sqrt{k_{2} - i k_{1}}
   f_{1}    = 0 \; ,
    $$

\noindent let us express it in terms of  new functions
$$
\sqrt{k_{2} - i k_{1}} f_{1} = e^{z} \varphi_{1} \; , \qquad \sqrt{k_{2}+ i k_{1}} f_{2} = e^{z} \varphi_{2} \,;
\eqno(7.1)
$$

\noindent
as a result we obtain
$$
\left (   Z {d \over d Z}  -i p  \right  )  \varphi_{1} +
 Z  \varphi_{2}    = 0 \; ,
 $$
 $$
\left (  Z {d \over d Z}  +ip  \right )  \varphi_{2}  +   Z \varphi_{1}    = 0 \; .
   \eqno(7.2)
$$

\noindent
Let us translate them to  the new variable
$
x = i Z = i \sqrt{k_{1}^{2} + k_{2}^{2}} e^{z}
$:
$$
\left (   x {d \over d x}  -i p  \right  )  \varphi_{1} -ix   \varphi_{2}    = 0 \; ,
$$
$$
\left (  x {d \over d x}  +ip  \right )  \varphi_{2}  -ix  \varphi_{1}    = 0 \; .
   \eqno(7.3)
$$

\noindent
From (7.3) it follow two  second-order equations
$$
( {d^{2} \over  d x^{2} }+  1+ {p^{2} + i p \over  x^{2}} ) \varphi_{1}=0 \;,
$$
$$
( {d^{2} \over  d x^{2}} +  1+ {p^{2} - i p \over  x^{2}} ) \varphi_{2}=0 \; .
\eqno(7.4)
$$

\noindent
Separating the factor  $\sqrt{x}:  \; \varphi_{1} =\sqrt{x} F_{1},\;  \varphi_{2} =\sqrt{x} F_{2}$,
we arrive at the two Bessel equations
$$
\left ( {d^{2} \over  d x^{2} }+  {1 \over x} {d \over dx } +  1+ {p^{2} + i p -1/4\over  x^{2}}\right  ) F_{1}=0\; ,
$$
$$
\left ( {d^{2} \over  d x^{2}} + {1 \over x} {d \over dx }  +  1 + {p^{2} - i p -1/4 \over  x^{2}} \right ) F_{2}=0\; .
\eqno(7.5)
$$

Since we are (especially) interested in  the case of the Weyl neutrino
(with helicity $-1$), we will continue to consider in detail  the case of
negative $p = - \sqrt{\epsilon^{2} - m^{2}}$. To avoid  new notation,
we will make a minus sign in front of $p$ in (7.5), so that
$$
\left ( {d^{2} \over  d x^{2} }+  {1 \over x} {d \over dx } + 1 + {p^{2} - i p -1/4\over  x^{2}}\right  ) F_{1}=0\; ,
$$
$$
\left ( {d^{2} \over  d x^{2}} + {1 \over x} {d \over dx }  +  1+ {p^{2} + i p -1/4 \over  x^{2}} \right ) F_{2}=0\; ,
\eqno(7.6)
$$

\noindent where $p =+\sqrt{\epsilon^{2} - m^{2}}>0$.

Thus, for polarization states with $p =+\sqrt{\epsilon^{2} - m^{2}}>0$,
the functions $F_ {1} (y),F_ {2} (y) $ satisfy the Bessel equations \cite{Kratser-1963}
$$
\left (  {d^{2} \over d x^{2}}  +  {1 \over x} {d \over d x} +1  -  {(ip+1/2)^{2} \over  x^{2} }  \right ) F_{1}=0 \; ,
$$
$$
\nu = -ip -1/2\;, \qquad
F_{1} =  J_{+\nu}(x) , \;  J_{-\nu}(x) \; .
\eqno(7.7a)
$$

$$
\left (  {d^{2} \over d x^{2}}  +  {1 \over x} {d \over d x} +1  -  {(-ip+1/2)^{2} \over  x^{2} }  \right ) F_{2}=0 \; ,
$$
$$
\mu = -ip +1/2=\nu+1 \;, \qquad
F_{2} =  J_{+\mu}(x) , \;  J_{-\mu}(x) \; .
\eqno(7.7b)
$$

Transition to polarization states with
$p' =-\sqrt{\epsilon^{2} - m^{2}}>0$ is achieved  by replacement  in (7.7) positive  $p$ to negative $p'=-p$,
which results in

$$
\nu' = +ip -1/2= -\mu\;,  \qquad  \mu' = +ip +1/2=\nu'+1 = - \nu \; .
\eqno(7.7c)
$$

Since we have to follow  explicit form of both functions $ F_ {1}, F_ {2} $ (up to a relative factor),
let us return to the first order  equations (now in  the variable $ x = iZ$);
with notation  $\nu =-ip -1/2$ it reads
$$
 \left ( x{   d  \over  d  x} - \nu   \right )
F_{1}  =  -ix   \, F_{2} \, ,
\eqno(7.8a)
$$
$$
 \left  ( x {
d  \over  d  x} + \nu +1 \right   ) F_{2} =  -ix
F_{1}  \,. \eqno(7.8b)
$$

 From the above it is known that  functions $F_ {1} (x), F_ {2} (x)$ satisfy the Bessel equations (7.6).
 Let us  recall the well-known recurrence formulas for the solutions of Bessel equation  \cite{Kratser-1963} --
 write them in a convenient form
$$
\left (  x {d \over dx }  -  \nu  \right ) F _{\nu} (x)  = -xF_{\nu +1} (x) \; ,
\eqno(7.9a)
$$
$$
\left ( x {d \over  dx }   - \nu  \right ) F _{-\nu} (x) = + x F_{-\nu -1  } (x)  \; ;
\eqno(7.9b)
$$

\noindent
here by $F_{\pm \nu}$ can be understood either of the Bessel functions:  $J_ {\pm \nu}$,
or Hankel functions  $H ^ {1} _ {\pm \nu}, H ^ {2} _ {\pm \nu}$, or Neumann functions
$N_{\pm \nu}(x)$.

Comparing eq. $(7.9a)$ with  eq.  of $(7.8a)$, we find two types of solutions:

\vspace{3mm}
in Bessel's functions
$$
I \qquad F_{1}^{I} (x) = J_{+\nu}(x)\;, \qquad  F_{2}^{I}(x) =  -i \; J_{+(\nu +1) }(x) \; ;
$$
$$
II  \qquad F_{1}^{II}(x)  = J_{-\nu}(x)\;, \qquad  F_{2} ^{II}(x)  =  +i \; J_{-(\nu +1) }(x) \; ;
$$
$$
\eqno(7.10)
$$

in Hankel's functions

$$
I \qquad F_{1}^{I} (x) = H^{1}_{+\nu}(x)\;, \qquad  F_{2}^{I}(x) =  -i \; H^{1}_{+(\nu +1) }(x) \; ;
$$
$$
II \qquad F_{1}^{II} (x) = H^{2}_{+\nu}(x)\;, \qquad  F_{2}^{II}(x) =  -i \; H^{2}_{+(\nu +1) }(x) \; ;
$$
$$
\eqno(7.11a)
$$

$$
I'  \qquad F_{1}(x)  = H^{1}_{-\nu}(x)\;, \qquad  F_{2} (x)  =  +i \; H^{1}_{-(\nu +1) }(x) \; ;
$$
$$
II'  \qquad F_{1}(x)  = H^{2}_{-\nu}(x)\;, \qquad  F_{2} (x)  =  +i \; H^{2}_{-(\nu +1) }(x) \; ;
$$
$$
\eqno(7.11b)
$$

\noindent note that $H^{1}_{-\nu}(x) = e^{i\nu \pi} H^{2}_{\nu}(x)$,
so the primed cases $I', II'$ coincide respectively with  $II,
I$  and by this reason will not be considered below.

And in Neumann functions
$$
I \qquad F_{1}^{I} (x) = N_{+\nu}(x)\;, \qquad  F_{2}^{I}(x) =  -i \; N_{+(\nu +1) }(x) \; ;
$$
$$
II  \qquad F_{1}^{II}(x)  = N_{-\nu}(x)\;, \qquad  F_{2} ^{II}(x)  =  +i \; N_{-(\nu +1) }(x) \; .
\eqno(7.12)
$$

\vspace{3mm}

First, let us detail  solutions in Bessel's functions \cite{Kratser-1963}.

In the region
$
z \rightarrow - \infty,\; x \rightarrow i 0\;,
$
$$
I \qquad F_{1}^{I} (x) = J_{+\nu}(x) = ({x \over 2})^{\nu} = (i\lambda )^{\nu} \;  e^{-ipz} e^{-z/2  } \; ,
$$
$$
\qquad \qquad  \;\; F_{2}^{I}(x) =  -i \; J_{+(\nu +1) }(x) =-i  \; (i\lambda )^{\nu+1 } \;  e^{-ipz} e^{+z/2} \; ,
$$

$$
II  \qquad  \;\; F_{1}^{II}(x)  = J_{-\nu}(x) = ({x \over 2})^{-\nu}  =  (i\lambda )^{-\nu} \; e^{+ipz} e^{ + z/2  } \;,
$$
$$
\qquad  \qquad F_{2} ^{II}(x)  =  +i \; J_{-(\nu +1) }(x)  = +i \; (i\lambda )^{-\nu-1} \; e^{+ipz} e^{ - z/2  } \; ;
$$
$$
 \eqno(7.13)
$$

\noindent
here and below we use the notation
$$
\lambda = {\sqrt{k_{1}^{2} + k_{2}^{2}} \over 2} \; .
$$

\noindent
In the region
$
z \rightarrow + \infty,\; x \rightarrow  +\infty  i$,
using the knows asymptotic formula \cite{Kratser-1963}
$$
J_{\nu}(x) \sim \sqrt{{2 \over \pi x}} \; \cos \left  ( x - (\nu +{1\over 2}) {\pi\over 2} \right ) ,
$$

\noindent
we get
$$
J_{+\nu}(z \rightarrow \infty ) \sim \sqrt{{2 \over i \pi X}} \; \cos  i \left  ( X  +  {p\pi\over 2} \right )
\longrightarrow \; \sqrt{{1 \over 2 \pi i  X}} \;e^{+p \pi/2} \;  e^{X}\; ,
$$
$$
J_{-\nu}(z \rightarrow \infty ) \sim \sqrt{{2 \over i \pi X}} \; \sin  i\left (  X  - {p \pi\over 2} \right )
\longrightarrow \;  i \sqrt{{1 \over 2 \pi i  X}} \;e^{- p \pi/2} \; e^{X} \; .
$$
$$
\eqno(7.14)
$$

Thus, the solutions (7.10) behave

\vspace{3mm}
$
z \rightarrow + \infty,\; x   = i X  \rightarrow  +i \infty  $,
$$
I \qquad F_{1}^{I} (x) = J_{+\nu}(x) \sim  \sqrt{{1 \over 2 \pi i  X}}  \;e^{+p \pi/2}  \; e^{X}\;,
$$
$$
\qquad  F_{2}^{I}(x) =  -i \; J_{+(\nu +1) }(x)  \sim  +\sqrt{{1 \over 2 \pi i  X}} \;e^{+p \pi/2} \; e^{X}\; ;
$$

$$
II  \qquad F_{1}^{II}(x)  = J_{-\nu}(x) \sim i \sqrt{{1 \over 2 \pi i  X}}\;e^{- p \pi/2}  \; e^{X}
\;,
$$
$$
\qquad  F_{2} ^{II}(x)  =  +i \; J_{-(\nu +1) }(x)  \sim  - \sqrt{{1 \over 2 \pi i  X}} \;e^{- p \pi/2}  \; e^{X}\; .
\eqno(7.15)
$$

Let us consider solutions $(7.11a)$ in Hankel's functions \cite{Kratser-1963}. They are  determined   in terms of
$J_{\pm \nu}(x)$ as follows
$$
H^{1}_{\nu} (x)  = + {i \over \sin \nu \pi } \left ( e^{-i \nu \pi} J_{+\nu}(x) -  J_{-\nu}(x) \right  )
\; ,
$$
$$
H^{2}_{\nu} (x) =- {i \over \sin \nu \pi  } \left ( e^{+i \nu \pi} J_{+\nu}(x) -  J_{-\nu}(x) \right  )\; .
\eqno(7.16)
$$

From here it is easy to set behavior of the Hankel functions for small $x$:

\vspace{2mm}

$
z \rightarrow - \infty,\; x \rightarrow i 0\;,
$

$$
H^{1}_{\nu} (x)  \sim
+ {i \over \sin \nu \pi }\left ( e^{-i \nu \pi}   (i\lambda )^{\nu}  e^{-ipz} e^{-z/2 } -
 (i\lambda )^{-\nu} e^{ipz} e^{+z/2 }\right  )
$$
$$
\sim
 {i   \over \sin \nu \pi }\; e^{-i \nu \pi} \; (i\lambda)^{\nu} \;    e^{-ipz} e^{-z/2 }
\; ,
$$

$$
H^{2}_{\nu} (x) \sim -  {i \over \sin \nu  \pi } \left (  e^{+i \nu  \pi} (i\lambda)^{\nu}  e^{-ipz} e^{-z/2 }
-  (i\lambda )^{-\nu} e^{ipz} e^{+z/2 } \right  )
$$
$$
\sim   - { i \over \sin \nu  \pi }\;e^{+i \nu \pi}  (i\lambda )^{\nu} \;    e^{-ipz} e^{-z/2 } \; ,
$$
$$
\eqno(7.17a)
$$

$$
H^{1}_{\nu+1} (x)  \sim
+ {i \over \sin (\nu+1) \pi }\left ( e^{-i(\nu+1)\pi}  (i\lambda )^{\nu +1}  e^{-ipz} e^{+z/2 }  -
(i\lambda )^{-\nu-1}  e^{+ipz} e^{-z/2 }  \right  )
$$
$$
\sim
 { - i \over \sin (\nu+1)\pi }\;    (i\lambda )^{-\nu-1} \; e^{+ipz} e^{-z/2 }  \; ,
$$

$$
H^{2}_{\nu+1 } (x) \sim -  {i \over \sin (\nu+1)\pi } \left (  e^{i(\nu +1) \pi} (i\lambda )^{\nu +1}  e^{-ipz} e^{+z/2 }
-  (i\lambda )^{-\nu-1}  e^{+ipz} e^{-z/2 }  \right  )
$$
$$
\sim   { i  \over \sin (\nu +1)\pi }\; (i\lambda )^{-\nu-1} \; e^{+ipz} e^{-z/2 } \; .
$$
$$
\eqno(7.17b)
$$

Therefore the solution of the system in Hankel functions behave as follows

\vspace{2mm}
$
z \rightarrow - \infty,\; x \rightarrow i 0\;,
$
$$
I \qquad F_{1}^{I} (x) = H^{1}_{+\nu}(x) \sim
{i   \over \sin \nu \pi } e^{-i \nu \pi}  (i\lambda )^{\nu}     e^{-ipz} e^{-z/2 } \;,
$$
$$
\qquad \qquad \qquad   F_{2}^{I}(x) =  -i  H^{1}_{+(\nu +1) }(x) \sim
{ - 1 \over \sin (\nu+1)\pi }    (i\lambda )^{-\nu-1}  e^{+ipz} e^{-z/2 }
\; ;
$$

$$
II \qquad F_{1}^{II} (x) = H^{2}_{+\nu}(x) \sim
 - { i \over \sin \nu  \pi }e^{+i \nu \pi} \; (i\lambda )^{\nu}     e^{-ipz} e^{-z/2 } \;,
$$
$$
\qquad \qquad  \qquad  F_{2}^{II}(x) =
 -i  H^{2}_{+(\nu +1) }(x) \sim    { 1  \over \sin (\nu +1)\pi } (i\lambda )^{-\nu-1}  e^{+ipz} e^{-z/2 } \; .
 \eqno(7.18)
$$

Behavior of the Hankel functions of $ z \rightarrow + \infty $ is given by \cite{Kratser-1963}
$$
H^{1}_{\nu} (x) \sim \sqrt{{2 \over \pi x}} \; \exp \left [ + i \left
(x - {\pi \over 2} (\nu +{1\over 2}) \right ) \right ] ,
$$
$$
H^{2}_{\nu} (x) \sim \sqrt{{2 \over \pi x}} \; \exp
\left [ -i \left (x - {\pi \over 2} (\nu +{1\over 2}) \right ) \right ] ,
\eqno(7.19)
$$

\noindent
so that

\vspace{3mm}

$
z \rightarrow + \infty,\; x  =  i X\rightarrow  + \infty  i\;,
$
$$
H^{1}_{\nu} (x) \sim
 \sqrt{{2 \over i \pi X}} \; \exp \left [ + i \left
(i X - {\pi \over 2} (\nu +{1\over 2}) \right ) \right ] \sim
 \sqrt{{2 \over i \pi X}} \; e^{ -   p \pi /2 } \; e^{-X} \; ,
$$
$$
H^{2}_{\nu} (x) \sim
 \sqrt{{2 \over i \pi X}} \; \exp \left [ -i \left
 (iX - {\pi \over 2} (\nu +{1\over 2}) \right ) \right ]  \sim
$$
$$
\sim
\sqrt{{2 \over i \pi X}} \; \; e^{ +   p \pi /2 } \; e^{+X} \;,
 $$
 $$
H^{1}_{\nu+1} (x) \sim
 \sqrt{{2 \over i \pi X}} \; \exp \left [ + i \left
(i X - {\pi \over 2} (\nu +1 +{1\over 2}) \right ) \right ] \sim
$$
$$
\sim   -i\;
 \sqrt{{2 \over i \pi X}} \; e^{ -   p \pi /2 } \; e^{-X} \; ,
$$
$$
H^{2}_{\nu+1} (x) \sim
 \sqrt{{2 \over i \pi X}} \; \exp \left [ -i \left
 (iX - {\pi \over 2} (\nu +1  +{1\over 2}) \right ) \right ]   \sim
 $$
 $$
 \sim +i\;
 \sqrt{{2 \over i \pi X}} \; \; e^{ +   p \pi /2 } \; e^{+X} \; .
$$
$$
 \eqno(7.20)
 $$

$$
I \qquad F_{1}^{I} (x) = H^{1}_{+\nu}(x) \sim
\sqrt{{2 \over i \pi X}} \; e^{ -   p \pi /2 } \; e^{-X} \;,
$$
$$
\qquad \qquad \qquad  F_{2}^{I}(x) =  -i \; H^{1}_{+(\nu +1) }(x)  \sim  - \sqrt{{2 \over i \pi X}} \; e^{ -   p \pi /2 } \; e^{-X}
\; ;
$$
$$
II \qquad F_{1}^{II} (x) = H^{2}_{+\nu}(x) \sim
\sqrt{{2 \over i \pi X}} \; \; e^{ +   p \pi /2 } \; e^{+X} \;,
$$
$$
\qquad \qquad \qquad  F_{2}^{II}(x) =  -i \; H^{2}_{+(\nu +1) }(x) \sim
  \sqrt{{2 \over i \pi X}} \; \; e^{ +   p \pi /2 } \; e^{+X} \; .
\eqno(7.21)
$$

Finally, let us consider solutions in terms of the Neumann functions.
These functions are as follows expressed in terms of Bessel functions \cite{Kratser-1963}
$$
N_{\nu}(x) = {  \cos \nu \pi \; J_{\nu} (x) - J_{-\nu}(x)  \over \sin \nu \pi } \; ,
$$
$$
N_{-\nu}(x) = {  J_{\nu} (x) - \cos \nu \pi \;  J_{-\nu}(x)  \over \sin \nu \pi } \,.
\eqno(7.22)
$$

In the region $z \rightarrow + \infty, \; (x \rightarrow + \infty i)$ we use the asymptotic formulas
$$
N_{\nu}(x) \sim
 \sqrt{{2 \over i \pi X}} \; \sin \left  ( i X - (\nu +{1\over 2}) {\pi\over 2} \right )  ,
 $$
 $$
 \sin (iX - a) = {e^{-X -ia} - e^{+X+ia} \over 2 } \sim {i\over 2} e^{X} e^{+ia} \; ,
$$
\noindent
we get
$$
N_{+\nu}(z \rightarrow \infty ) \sim  + i\sqrt{{1 \over 2 \pi i  X}}  e^{p \pi/2} \; e^{X}\; ,
$$
$$
N_{+\nu+1}(z \rightarrow \infty )  \sim  \sim  - \sqrt{{1 \over 2 \pi i  X}} e^{p \pi/2}  \; e^{X}\; ,
$$
$$
N_{-\nu}(z \rightarrow \infty ) \sim   \sqrt{{1 \over 2 \pi i  X}}  e^{-p \pi/2}   \; e^{X} \; ,
$$
$$
N_{-\nu-1}(z \rightarrow \infty ) \sim  i \sqrt{{1 \over 2 \pi i  X}}  e^{-p \pi/2}   \; e^{X} \; .
\eqno(7.23)
$$

Thus, the solution (7.12) in the functions of the Neumann behave at $ z \rightarrow + \infty $ as follows:
$$
I \qquad F_{1}^{I} (x) = N_{+\nu}(x) \sim
i\sqrt{{1 \over 2 \pi i  X}}  e^{p \pi/2} \; e^{X} \; ,
 $$
$$
\qquad \qquad \qquad  F_{2}^{I}(x) =
 -i \; N_{+(\nu +1) }(x)  \sim
  +i \sqrt{{1 \over 2 \pi i  X}} e^{p \pi/2}  \; e^{X} \; ;
$$

$$
II  \qquad F_{1}^{II}(x)  = N_{-\nu}(x) \sim
 \sqrt{{1 \over 2 \pi i  X}}  e^{-p \pi/2}   \; e^{X} \;,
$$
$$
\qquad
\qquad \qquad  F_{2} ^{II}(x)  =  +i \; N_{-(\nu +1) }(x)  \sim
 - \sqrt{{1 \over 2 \pi i  X}}  e^{-p \pi/2}   \; e^{X} \; .
\eqno(7.24)
$$

Now, using the definition (7.22), we find the behavior of the Neumann functions at $z\rightarrow -\infty$.
$$
N_{\nu}(x) = {  \cos \nu \pi \; J_{\nu} (x) - J_{-\nu}(x)  \over \sin \nu \pi }
$$
$$
\sim
{  \cos \nu \pi \;  (i\lambda )^{\nu} \;  e^{-ipz} e^{-z/2  } -  (i\lambda )^{-\nu} \;  e^{+ipz} e^{ + z/2  }
\over \sin \nu \pi }
=
$$
$$
\sim   {  \cos \nu \pi \;   \over \sin \nu \pi } \; (i\lambda )^{\nu} \;   e^{-ipz} e^{-z/2  } \; ,
$$

$$
N_{\nu+1 }(x) = {  \cos (\nu+1)  \pi \; J_{\nu+1 } (x) - J_{-\nu-1}(x)  \over \sin (\nu +1)\pi }
$$
$$
\sim
{  \cos (\nu+1)  \pi \; (i\lambda )^{\nu+1} \;   e^{-ipz} e^{+z/2  } - (i\lambda )^{-\nu-1} \;
 e^{+ipz} e^{ - z/2  } \over \sin (\nu+1) \pi }
$$
$$
\sim{  1  \over \sin (\nu +1) \pi } \;(i\lambda )^{-\nu-1} \;  e^{+ipz} e^{-z/2  } \; ,
$$

$$
N_{-\nu}(x) =   {  J_{\nu} (x) - \cos \nu \pi \;  J_{-\nu}(x)  \over \sin \nu \pi }
$$
$$
\sim
{   (i\lambda )^{\nu} \;  e^{-ipz} e^{-z/2  } -
\cos \nu \pi \; (i\lambda )^{-\nu} \;   e^{+ipz} e^{ + z/2  }  \over \sin \nu \pi }
$$
$$
\sim
{ 1    \over \sin \nu \pi } \; (i\lambda /2)^{\nu} \;  e^{-ipz} e^{-z/2  } \;  ,
$$

$$
N_{-\nu-1}(x) = {  J_{\nu+1} (x) - \cos (\nu+1) \pi \;  J_{-\nu-1}(x)  \over \sin (\nu +1)\pi }
$$
$$
\sim {  (i\lambda )^{\nu+1 } \;  e^{-ipz} e^{+z/2  } - \cos (\nu +1)  \pi \;
 (i\lambda )^{-\nu-1} \; e^{+ipz} e^{ - z/2  }  \over \sin (\nu+1)  \pi }
 $$
 $$
 \sim
{ - \cos (\nu +1)  \pi     \over \sin (\nu +1) \pi } \; (i\lambda )^{-\nu-1} \;  e^{+ipz} e^{-z/2  } \,.
$$

Consequently, the solution (7.12) in the functions of the system behave Neumann $ z \rightarrow - \infty $ as follows:
$
z \rightarrow - \infty,\; x \rightarrow i 0\;,
$

$$
I \qquad F_{1}^{I} (x) = N_{+\nu}(x) \sim
 {  \cos \nu \pi \;   \over \sin \nu \pi } \; (i\lambda )^{\nu} \; e^{-ipz} e^{-z/2  } \;,
$$
$$
\qquad  \qquad F_{2}^{I}(x) =  -i \; N_{+(\nu +1) }(x) \sim
 -i {  1  \over \sin (\nu +1) \pi } \; (i\lambda )^{-\nu-1} \; e^{+ipz} e^{-z/2  } \; ;
$$

$$
II  \qquad F_{1}^{II}(x)  = N_{-\nu}(x) \sim
  { 1    \over \sin \nu \pi } \; (i\lambda )^{\nu} \; e^{-ipz} e^{-z/2  } \;,
$$
$$
 \qquad \qquad\;\;\;
 F_{2} ^{II}(x)  =  +i \; N_{-(\nu +1) }(x)  \sim
+i { - \cos (\nu +1)  \pi     \over \sin (\nu +1) \pi } \; (i\lambda )^{-\nu-1} \; e^{+ipz} e^{-z/2  } \; .
$$
$$
\eqno(7.25)
$$

The results can be collected  into a table.

\vspace{2mm}

$\sigma = - p$

$$
\hspace{-5mm}
\left. \begin{array}{ccccccccc}
 & \mbox{Bessel}  &  \mbox{} &
 & \mbox{Hankel} &  \mbox{} &
 & \mbox{Neumann}  &  \mbox{}\\[2mm]
 & z \rightarrow -\infty &  z \rightarrow +\infty  &
 & z \rightarrow -\infty &  z \rightarrow +\infty &
 & z \rightarrow -\infty &  z \rightarrow +\infty \\[2mm]
Z_{1}^{I} & e^{-ipz} &  e^{+X}  &
Z_{1}^{I} & e^{-ipz} &  e^{-X}  &
Z_{1}^{I} & e^{-ipz} &  e^{+X}\\
Z_{2}^{I} & 0  &  e^{+X}  &
Z_{2}^{I} & e^{+ipz} &  e^{-X}  &
Z_{2}^{I} & e^{+ipz} &  e^{+X}  \\
Z_{1}^{II} & 0  &  e^{+X}  &
Z_{1}^{II} & e^{-ipz} &  e^{+X}  &
Z_{1}^{II} & e^{-ipz} &  e^{+X}\\
Z_{2}^{II} & e^{+i[z}  &  e^{+X}  &
Z_{2}^{II} & e^{+ipz} &  e^{+X}  &
Z_{2}^{II} & e^{+ipz} &  e^{+X}
\end{array} \right.
$$

The most interesting are solutions of the type $I$  in Hankel functions:

$$
z \rightarrow - \infty \;, \qquad Z^{I}_{1} \sim e^{-ipz} \; , \qquad Z^{I}_{2} \sim e^{+ipz} \; ,
$$
$$
z \rightarrow + \infty \;, \qquad Z^{I}_{1} \sim e^{-X} \rightarrow 0 \; , \qquad Z^{I}_{2} \sim e^{-X}
\rightarrow 0 \; ,
$$

\noindent which means that  the problem posed in
Lobachevsky space can simulate a situation in the flat space for a
quantum-mechanical particle  of spin 1/2  (Dirac electron and Weyl neutrino)
in a 2-dimensional
 potential barrier smoothly rising to infinity on the right.
Electromagnetic field  behaves itself similarly -- see in \cite{Ovsiyuk-Red'kov-2011}.

\subsection*{Conclusions and acknowledgement}

In the paper  complete systems of exact solutions for Dirac and
Weyl equations in the Lobachevsky space $H_{3}$ are constructed on
the base of the method of separation of the variables in
quasi-cartezian coordinates.An  extended  helicity  operator
is intoduced. It is shown that solution constructed when translating to the limit of vanishing curvature
coincide with common plane wave solutions on Minkowski space going in opposite $z$-directions.
Electromagnetic field  behaves itself similarly.

Author  is grateful  to  V.M. Red'kov for  encouragement  and
advices.

\end{document}